\documentclass[a4paper]{jpconf}
\usepackage{graphicx}
\usepackage{amsmath,amssymb,graphicx} 
\usepackage{subfigure}
\usepackage{amsfonts}
\usepackage[english]{babel}
\usepackage[applemac]{inputenc}
\usepackage{graphicx}
\usepackage{cancel}
\usepackage{color}
\usepackage{hyperref}
\usepackage{afterpage}

\begin{document}

\title{Winding angle distributions for two-dimensional collapsing polymers}

\author{Arturo Narros}
\address{School of Mathematical Sciences, Queen Mary University of London, Mile End Road
E1 4NS London, UK}
\author{Aleksander L Owczarek}
\address{School of Mathematics and Statistics, The University of Melbourne, Parkville, Vic 3010, Australia}
\author{Thomas Prellberg}
\address{School of Mathematical Sciences, Queen Mary University of London, Mile End Road
E1 4NS London, UK}

\ead{a.n.gonzalez@qmul.ac.uk,owczarek@unimelb.edu.au,t.prellberg@qmul.ac.uk}

\date{\today}

\begin{abstract}
We provide numerical support for a long-standing prediction of universal scaling of winding angle distributions.    Simulations of interacting self-avoiding walks show that the winding angle distribution for $N$-step walks is compatible with the theoretical prediction of a Gaussian with a variance growing asymptotically as $C\log N$, with $C=2$ in the swollen phase (previously verified), and $C=24/7$ at the $\theta$-point. At low temperatures weaker evidence demonstrates compatibility with the same scaling and a value of $C=4$ in the collapsed phase, also as theoretically predicted.
\end{abstract} 

\section{Introduction}

Polymers in a dilute solution can be either swollen or collapsed, or at an intermediate critical point, the so-called $\theta$-point, depending on the quality of the solvent \cite{gennes1979a-a}. The canonical lattice model for polymer collapse is the interacting self-avoiding walk on a regular lattice such as the square or simple cubic lattice, in two or three spatial dimensions, respectively.

Some time ago, the winding angle distribution for polymers received theoretical attention. In two dimensions, the winding angle distribution of Brownian paths of length $N$ satisfies Spitzer's law \cite{spitzer1958a-a}
\begin{equation}
P(x=2\theta/\log N)\sim\frac1{1+x^2}\;.
\end{equation}
This distribution has divergent moments due to the infinitesimal winding centre. Changing to macroscopic winding centres, or, alternatively, to random walks on a lattice, changes the structure of the winding centre and leads to an exponential decay in the winding angle distribution. More precisely, the distribution gets altered to
\begin{equation}
P(x=2\theta/\log N)\sim\frac1{\cosh^2(1+\pi x/2)}
\end{equation}
for adsorbing winding centres and to
\begin{equation}
P(x=2\theta/\log N)\sim\frac1{\cosh(1+\pi x/2)}
\end{equation}
for reflecting winding centres \cite{rudnick1987a-a,belisle1989a-a,saleur1994a-a}. 

The introduction of an excluded volume leads to a swelling of the polymer. Heuristically, swelling implies that the winding of segments $N/2$, $N/4$, $N/8$, $\ldots$ become approximately independent. Hence an $N$-step chain has $O(\log N)$ segments and a law-of-large-numbers argument implies that the winding angle distribution becomes Gaussian with variance proportional to $\log N$. More precisely, one finds that
\begin{equation}
P(x=\theta/\sqrt{\log N})\sim\exp(-x^2/(2C)).
\end{equation}
Note that the introduction of the excluded volume has changed the scaling variable from $x=\theta/\log N$ to $x=\theta/\sqrt{\log N}$. 

Using a mapping to a Coulomb gas \cite{duplantier1988a-a}, it has been argued that the constant $C$ is universal in two dimensions, in that it assumes distinct values depending only on the phase the polymer is in. In particular, it has been predicted that
\begin{equation}
C=\begin{cases}2&\mbox{swollen phase,}\\ 24/7&\mbox{critical state,}\\ 4&\mbox{dense phase.}\end{cases}
\end{equation}
This calculation has been extended to watermelon configurations in \cite{prellberg1998a-a}, where it was shown that the number $L$ of legs in the watermelon configuration modifies the constant $C$ by a factor $1/L^2$. As a consequence, it was argued that for $\theta$-point polymers on the Manhattan lattice one finds $C=6/7$, as the modification of the lattice leads to a choice of $L=2$ instead of $L=1$. The reason for this is that at the $\theta$-point the polymer configuration forms part of a percolation cluster hull, and thus the presence of a virtual second polymer completing the hull is felt. This argument has been confirmed by simulations \cite{prellberg1998b-a}.

While the predicted values of $C$ in the swollen phase and at the Manhattan lattice $\theta$-point have been confirmed by simulations, there has been no confirmation of the predicted value $C=24/7$ at the $\theta$-point. There has also been some argument in the past whether the collapsed phase can indeed be identified with a dense phase. Moreover, the heuristic argument leading to a law-of-large-numbers could break down for the collapsed phase, as the chain is no longer swollen, so that a numerical verification of the predicted value $C=4$ is especially important.

We note that there have also been simulations of interacting self-avoiding walks up to length $N=300$ \cite{chang2000a-a}, suggesting that the results at the $\theta$-point and in the collapsed phase are more consistent with a stretched exponential of the type $\exp(-|\theta|^\alpha/C\log N)$ with $\alpha\approx1.5$. As a consequence, the scaling variable would change to $x=\theta/(\log N)^{1/\alpha}$. 

Here we show, using simulations of interacting self-avoiding walks up to $N=400$ with flatPERM, which allows simultaneous uniform sampling with respect to the number of interactions and the winding angle, that the winding angle distributions are indeed asymptotically Gaussian with a scaling variable $x=\theta/\sqrt{\log N}$, and that the variance grows linearly in $\log N$ with the value of $C$ being in agreement with the theoretical predictions for swollen, critical, and dense polymers.

\section{Simulation Details}
\label{sec:simdet}

A large ensemble of linear chain configurations was obtained on a square lattice by means 
of the flatPERM algorithm proposed by Prellberg and Krawczyk in \cite{prellberg2004a-a}, which is
an extension of the Pruned and Enriched Rosenbluth Method (PERM) proposed in \cite{grassberger1997a-a}. The algorithm is able to obtain statistical weights for a specific 
size and energy level ($N,e$). FlatPERM was successfully applied to several models in a lattice 
such as, for example, asymmetric interacting self-avoiding  trails (ISAT) \cite{bedini2013c-:a}.

The flatPERM (multi-parametric flatPERM), described here, is able to obtain 
the aforementioned weights as a function of 
chain size, energy and any additional observable considered($N,e,O$).
An advantage of multi-parametric flatPERM is  the estimation of micro-canonical 
partition function of any walker model as a function of
$N,e$ and the observable considered.
In addition, a suitable observable considered can enhance 
the calculation pulling out ``strange'' configurations, or rather less statistical 
probable configurations. 

A key variable in flatPERM is the ratio ($r$) between the the actual 
weight of a chain configuration of size $n$ and energy $m$, $W_{n,m}$, and
an estimation of it, $C_{n,m}^{est}$. If $r$ is bigger than one we enrich, otherwise 
we prune. This ratio is changed accordingly in our version as
\begin{equation}\label{sec:simdet:eq1}
r = \frac{W_{n,m,k}}{C_{n,m,k}^{est}}\,,
\end{equation}
where the $k$ index corresponds to the value of the 
observable $O$ in a configuration, and
\begin{equation}\label{sec:simdet:eq2}
C_{n,m,k}^{est} = < W >_{n,m,k} = \frac{1}{S} \sum_i W_{n,m,k}^{(i)}\,,
\end{equation}
i.e, the average of all the generated samples $i$ with size $n$, energy $m$ and 
value of the observable $k$.

The observable implemented in this work is the winding angle 
($w$)~\cite{rudnick1987a-a}, i.e., the angle about a rod or reference point
that is swept out by growing polymer or random walker.
Winding angle is well defined in two dimensions and 
has been studied extensively in the literature for several kinds of walkers
~\cite{saleur1994a-a,duplantier1988a-a,prellberg1998b-a,samokhin1998a-a,rudnick1988a-a}. 
Winding angle is an interesting observable, since it gives geometric 
information about the degree of rolling up about the origin.
Therefore, it is a excellent candidate to test our new implementation of flatPERM.

The flatPERM~\cite{prellberg2004a-a} algorithm has a technical problem
due to the inaccuracy of the initial weights. This problem is even harder in our version, since the
additional parameter can create degeneracy for each energy level $m$ (weight).
Degeneracy ($d$) produces a lot of initial enrichment, $r \gg 1$, because the estimated weight is
a partial average of the initial configuration with same weight (energy). Then the initial run of
the presented flatPERM version is characterized by a lot of enrichment, in order to obtain a
proper estimation of the weight. This problem is particularly difficult if the observable is not 
very related with energy (weight), and it is a continuous function that, as in our case, should be 
discretized. 

Histogram binning is a critical ingredient in this algorithm that will determine 
the degree of degeneracy, $d$. The smaller the binning the more enrichment that will be necessary. 
A priori an optimal binning will be one that reduces degeneracy, improves the generation of
``less usual'' walks, and ensures the corresponding histogram fits in the computer memory available.
The calculations here presented have used a general histogram binning of $10$ degrees. 
It is small enough to identify all the possible configurations for the smallest chains 
and the corresponding histogram fits in the computer memory available to us.
However, this binning might be not optimal for big chains, given the big
amount of winding angles possible.

Initial enrichment issues were solved integrating a certain percentage ($g$) 
of the degeneracy level around the location of the winding angle:
\begin{equation}\label{sec:simdet:eq3}
\hat{C}_{n,m,k} = \frac{1}{S} \sum_{i=k-a}^{k+a} \sum_j C_{n,m,i}^{(j)}	\,,
\end{equation}
where $a = d\cdot g/100$. 

The advantages of this procedure are several: (i)
it allows to get better initial estimations of the weights preventing 
too much enrichment; (ii) it can be equally applied to be energy to improve 
the issue above mentioned with initial wrong weights; 
(iii) it is size dependent and can be used with any kind of observable; 
(iv) the integration percentage, $g$, can be reduced along the run in order to increase 
the degeneracy (enrichment) and, therefore, the searching of ``unusual'' configurations.

On the other hand, this implementation suffers from two major disadvantages:
(i) parameter dependency due to the binning; and (ii) is computationally expensive in terms of
time and storage, given the large dimensional histogram that must be obtained.
However, the second disadvantage can be easily solved with parallel computing. 

Calculations here presented started with $g = 100$ \% for
$5\times10^5$ tours, that was subsequently reduced to 5 \%.
The total number of tours, $S$ in eq~\ref{sec:simdet:eq2} and~\ref{sec:simdet:eq3},
it is about $2\times10^6$. It is worth mentioning that our implementation 
allows restarting runs, saving the ful state of enrichment/prunning along the chain. 
Thus, it is possible combine several runs with different random seeds to decrease the 
correlations between samples mentioned in flatPERM~\cite{prellberg2004a-a}.

\section{Results}
\label{sec:res}

\begin{figure}[!hbt]
\begin{center}
\includegraphics[width=1\columnwidth]{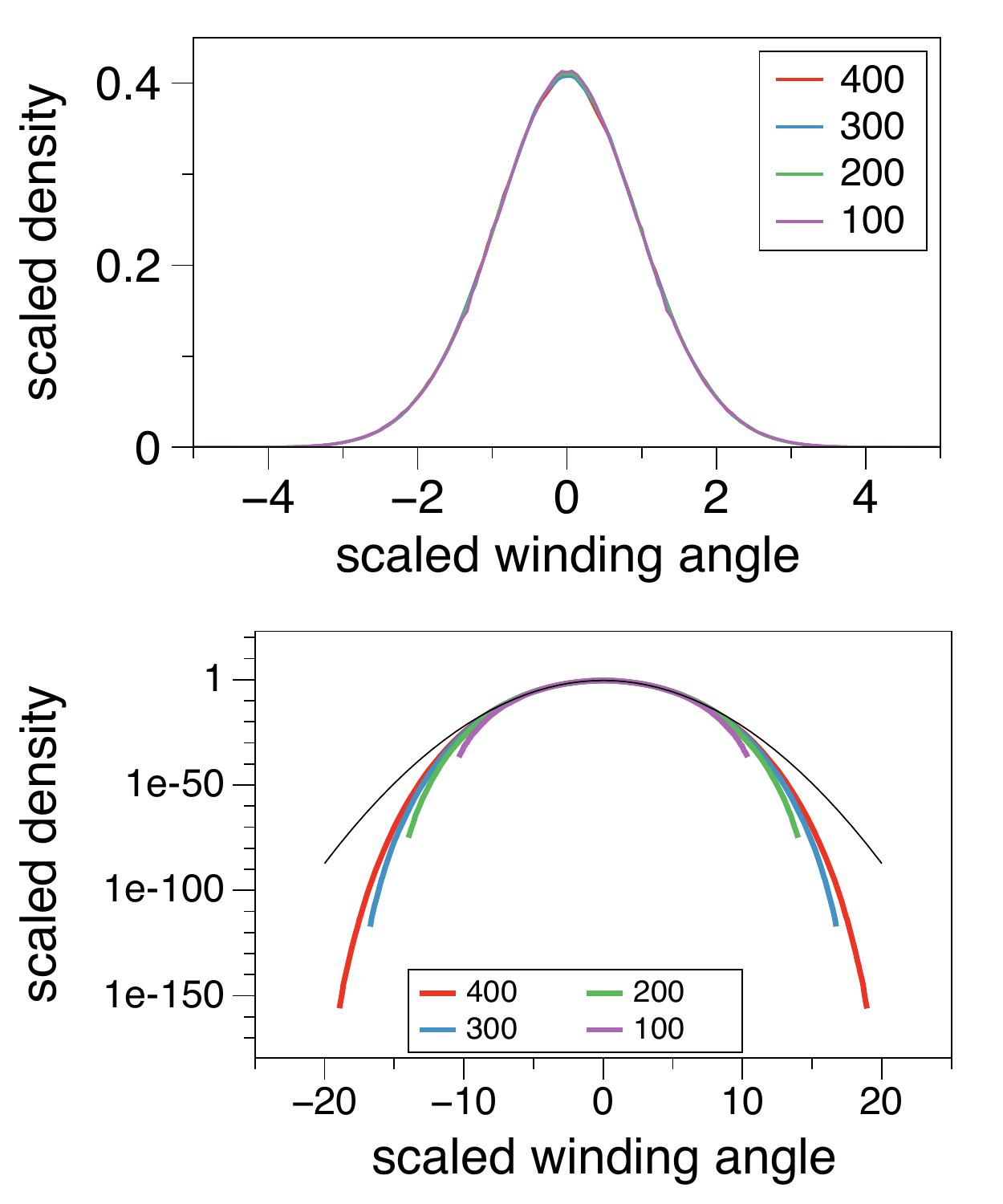}
\caption{Top: normalised winding angle distributions at $N=100$, $200$, $300$, and $400$ for $\beta=0$.
         Bottom: the same data in a semi-logarithmic plot, indicating the spread of the distribution as $N$ increases.
             The thin line indicates the normal distribution $\exp(-x^2/2)/\sqrt{2\pi}$.\vspace*{0.5cm}}
	\label{sec:res:fig1}
\end{center}
\end{figure}

\begin{figure}[!hbt]
\begin{center}
\includegraphics[width=1\columnwidth]{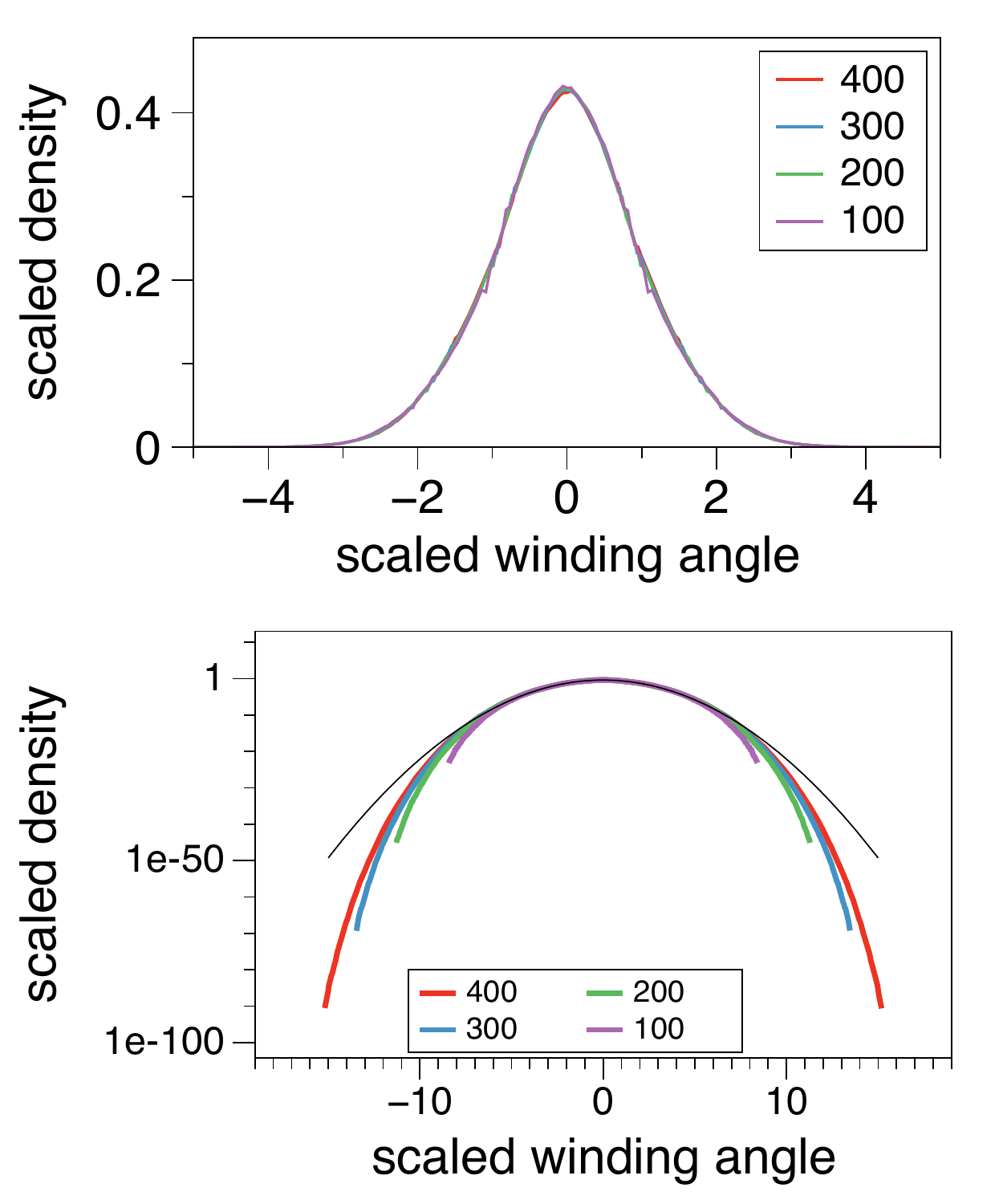}
\caption{Top: normalised winding angle distributions at $N=100$, $200$, $300$, and $400$ for $\beta=0.6673$.
         Bottom: the same data in a semi-logarithmic plot, indicating the spread of the distribution as $N$ increases.
             The thin line indicates the normal distribution $\exp(-x^2/2)/\sqrt{2\pi}$.\vspace*{0.5cm}}
	\label{sec:res:fig2}
\end{center}
\end{figure}

\begin{figure}[!hbt]
\begin{center}
\includegraphics[width=1\columnwidth]{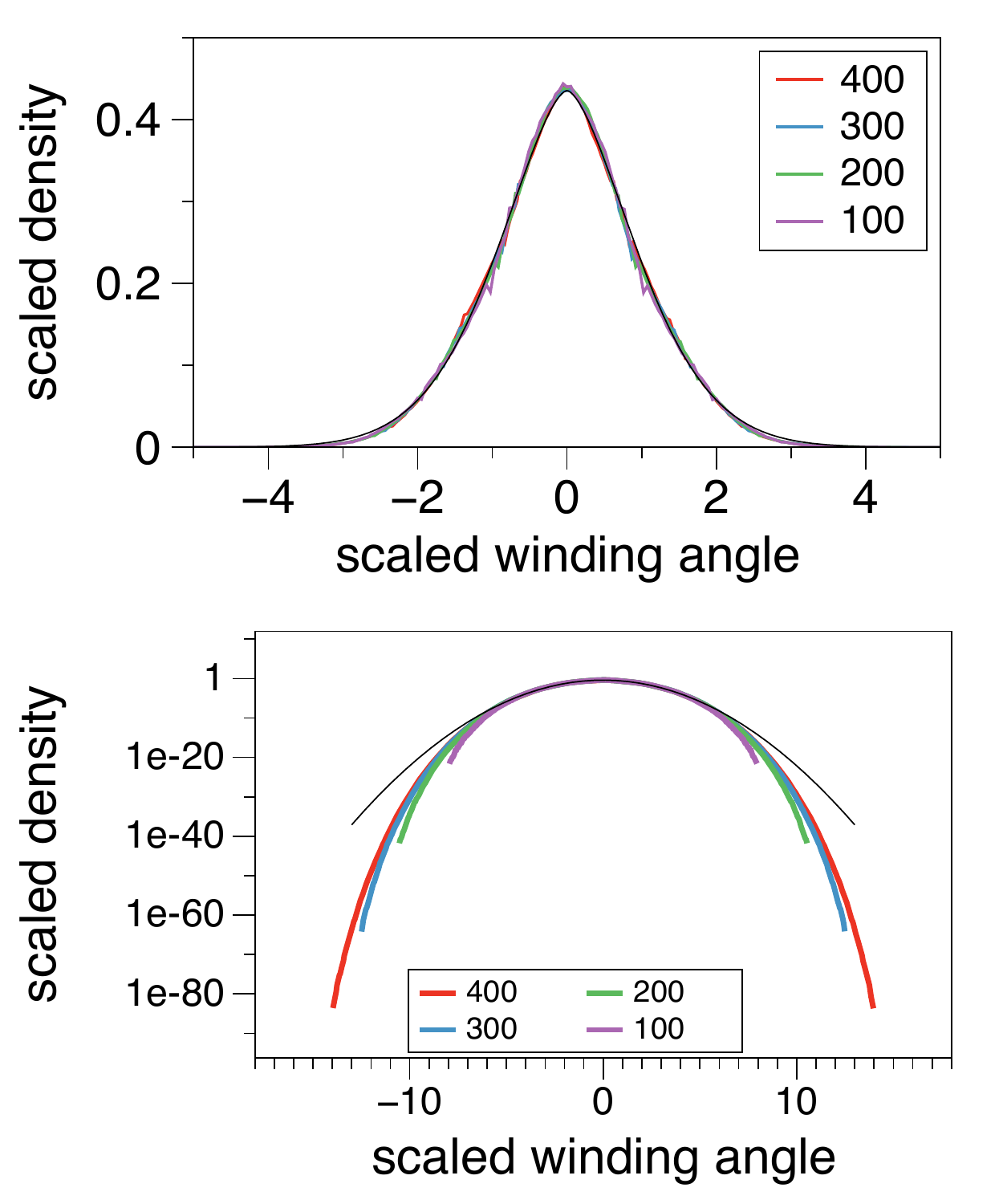}
\caption{Top: normalised winding angle distributions at $N=100$, $200$, $300$, and $400$ for $\beta=0.8$.
         Bottom: the same data in a semi-logarithmic plot, indicating the spread of the distribution as $N$ increases.
             The thin line indicates the normal distribution $\exp(-x^2/2)/\sqrt{2\pi}$.\vspace*{0.5cm}}
	\label{sec:res:fig3}
\end{center}
\end{figure}

From the two-parameter histograms in the number of interactions and winding angle obtained, we can now extract winding angle distributions at several temperatures. We have chosen $\beta=0$ for the swollen phase, $\beta=0.8$ for the collapsed phase, and $\beta=0.6673$ for the $\theta$-point, using the best published estimate $\beta_\theta=0.6673(5) $\cite{caracciolo2011a-a}. (We note in passing that previous work \cite{chang2000a-a} used values $0$, $\beta_\theta=0.658(4)$, and $0.8$.)

In Figure \ref{sec:res:fig1} we show winding angle distributions at infinite temperature ($\beta=0$) for sizes $N=100$ to $N=400$, scaled such that the distributions have unit variance. Uniform sampling with respect to the winding angle enables us to sample the distribution over more than 100 orders of magnitude. We see that the central part of the distribution approaches a Gaussian distribution very well. Convergence in the tails is much slower, as is evident from the semi-logarithmic plot. However, this is not surprising, as the full distribution is supported on an interval that grows as $\sqrt N$ due to the largest winding angle being obtained by tightly wound spirals.

In Figure \ref{sec:res:fig2} we show winding angle distributions at the $\theta$-point ($\beta=0.6673$) for sizes $N=100$ to $N=400$, again scaled to unit variance. The central part of the distribution again shows a nice scaling collapse. Note the significant change of the probability in the tail of the distribution, due to the fact that configurations with large winding also have large energy and thus have significantly higher probability at lower temperatures.

In Figure \ref{sec:res:fig3} we show scaled winding angle distributions in the low temperature regime ($\beta=0.8$) for sizes $N=100$ to $N=400$, again indicating scaling collapse.

\begin{figure}[!hbt]
\begin{center}
\includegraphics[width=0.6\columnwidth]{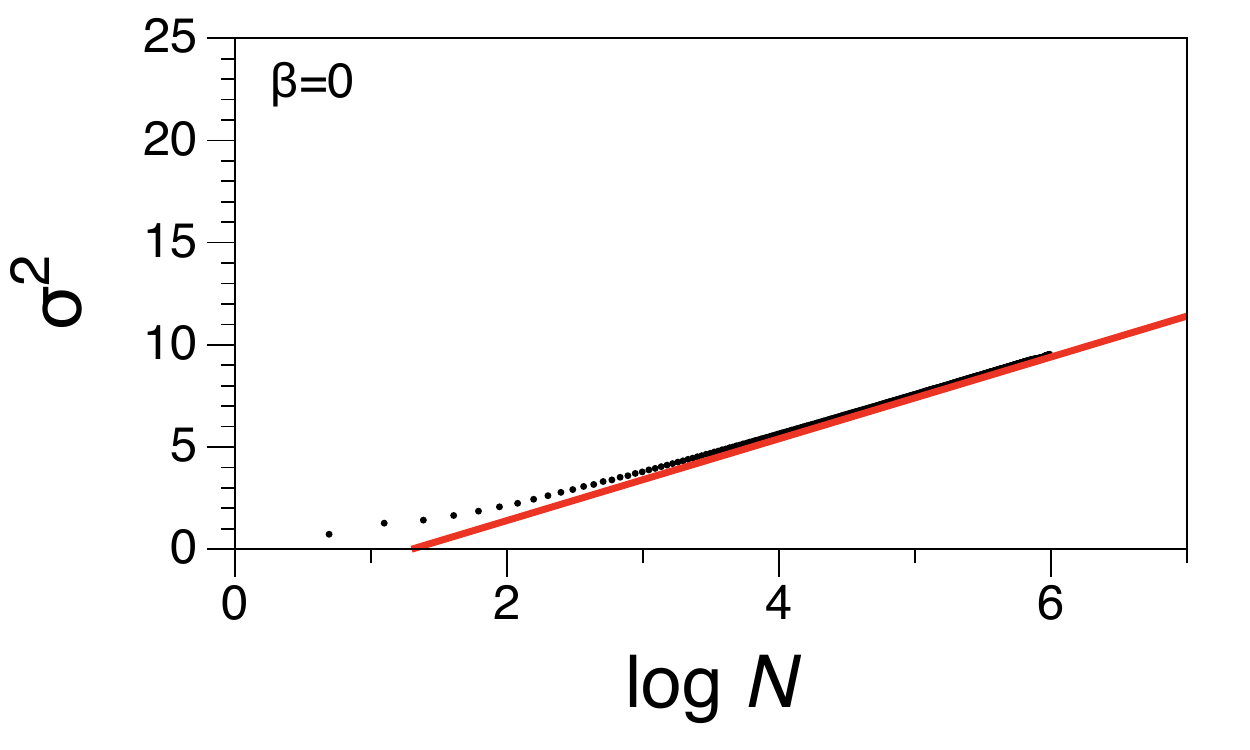}
\includegraphics[width=0.6\columnwidth]{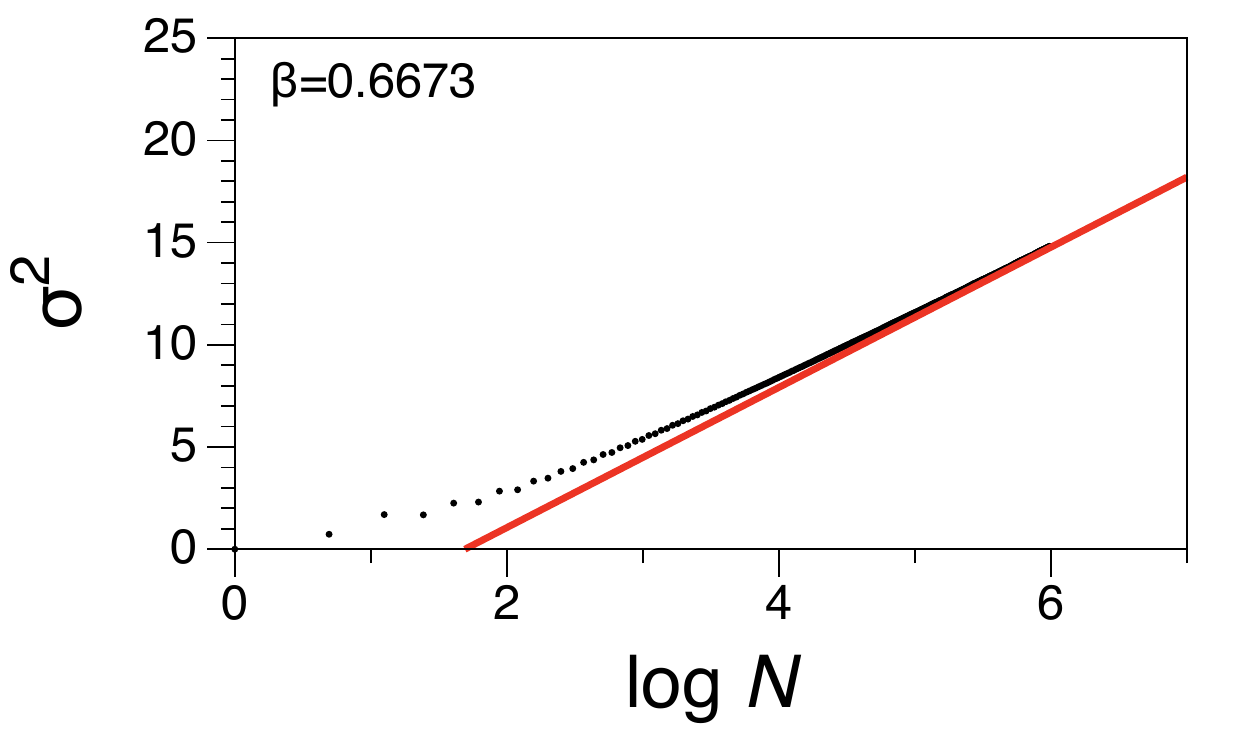}
\includegraphics[width=0.6\columnwidth]{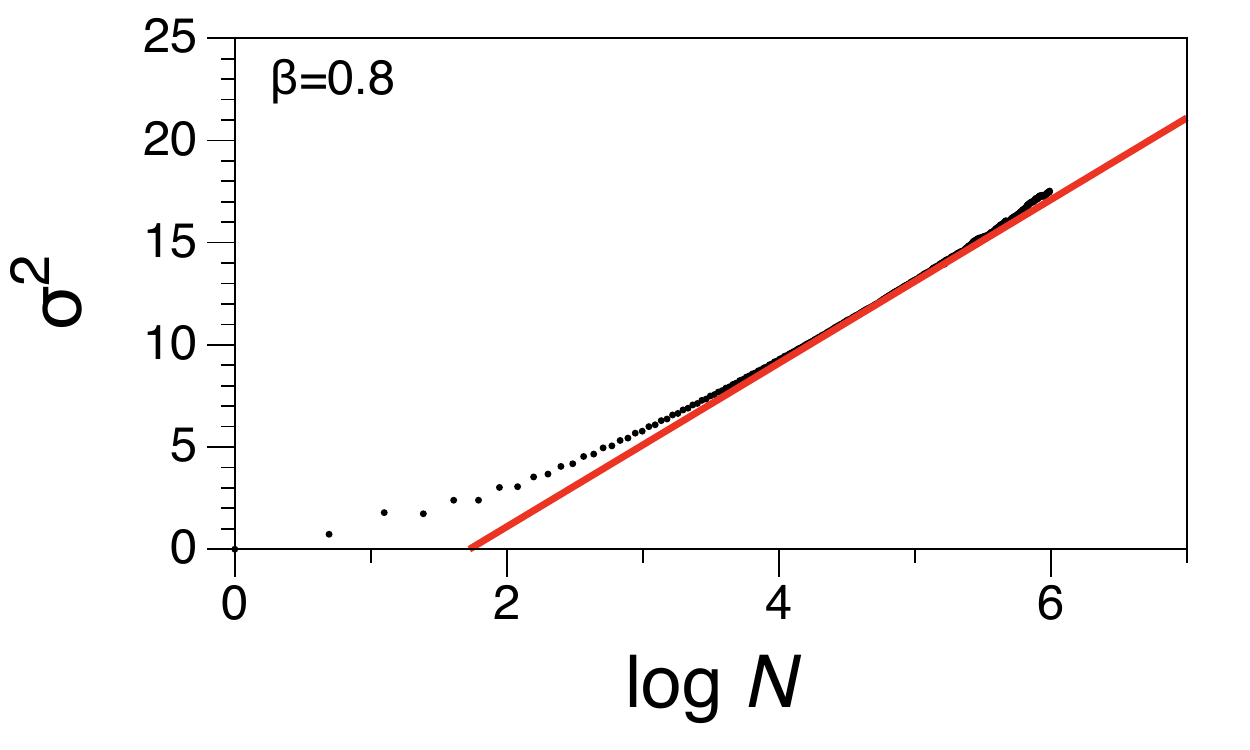}
\caption{Shown is the variance of the winding angle distribution for lengths up to $N=400$ (dots) at $\beta=0$ (top), $\beta=0.6673$ (middle), and  $\beta=0.8$ (bottom). The straight lines have slopes of 2 (top), $24/7$ (middle), and 4 (bottom) illustrating the compatibility of the data with the theoretical scaling predictions.There is curvature at short lengths, as one would expect coming from corrections to scaling. There is also some noisiness and perhaps curvature in the data for the longest lengths in the low temperature figure (bottom) when compared to the theoretical straight line.\vspace*{0.5cm}} \label{fig:figure4}
\end{center}
\end{figure}

\begin{figure}[!hbt]\begin{center}
\includegraphics[width=0.7\columnwidth]{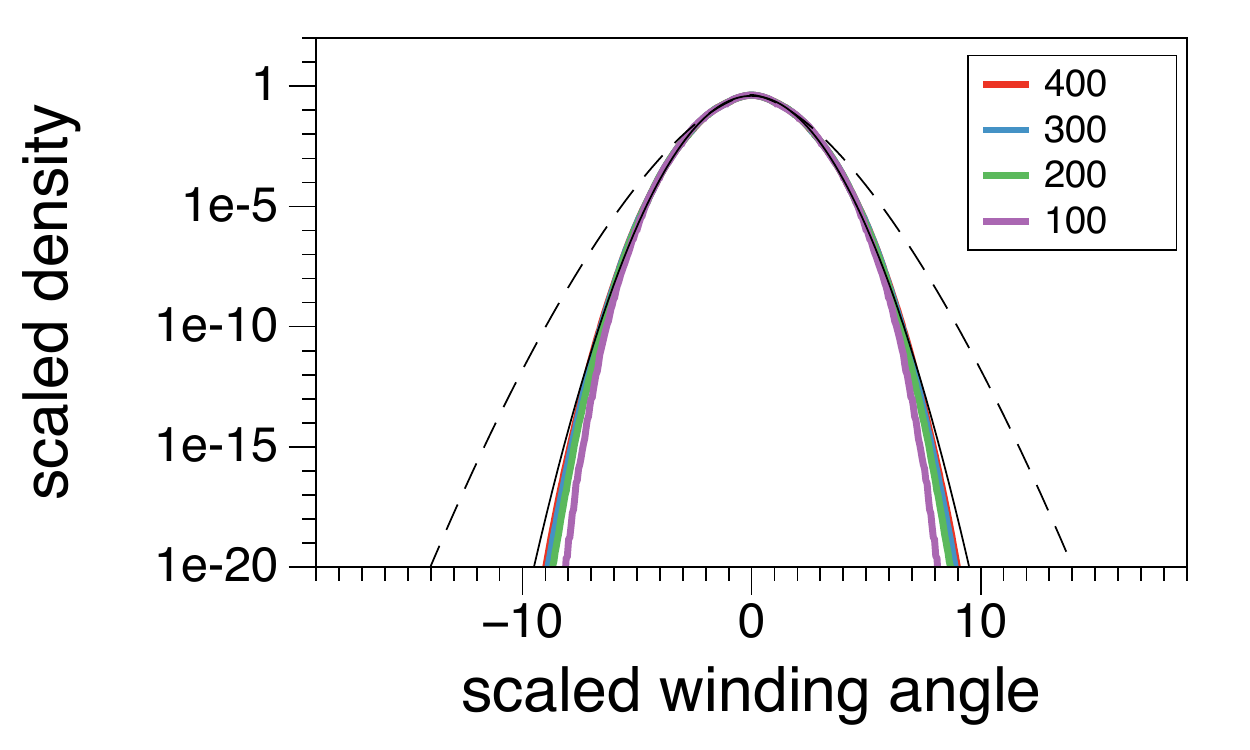}
\caption{Normalised winding angle distributions at $N=100$, $200$, $300$, and $400$ for $\beta=0.6673$.
             The thin line indicates the normal distribution $\exp(-x^2/2)/\sqrt{2\pi}$, whereas the dashed line indicates a stretched exponential with exponent $\alpha=1.6$.}
	\label{sec:res:fig7}
\end{center}
\end{figure}

Now that we have established that the central regions of the winding angle distributions collapse to the same distributions, we turn to the investigation of the growth of the winding angle variance in the different regimes. 
In Figure \ref{fig:figure4} we display the variance as a function of $\log N$. We also plot the three theoretical scaling predictions on these plots which are simply straight lines with slopes of the values of $C$ being 2, $24/7$ and $4$ in the different regimes, There is clearly some curvature in the data coming from finite-size effects. However we can confidentially infer compatibility at both high temperatures ($\beta=0$) and at the $\theta$-point ($\beta=0.6673$). At our low temperature value ($\beta=0.8$) we make a more qualified statement as there are larger errors in our data and still some curvature at longer lengths, so state that the theoretical prediction of 4 for $C$ is reasonable compatible with the data. In any case  there is clear disagreement with the $\theta$-point and low-temperature values of $2.75$ and $3.14$ given in \cite{chang2000a-a}.

Note also that the behaviour in the tails of the distribution does not indicate the presence of a stretched exponential. At the $\theta$-point, Figure \ref{sec:res:fig7} shows that while it is entirely plausible that there is convergence to a normal distribution, convergence to a stretched exponential, which has much fatter tails, seems quite unlikely. The figure shows a stretched exponential with exponent $\alpha=1.6$, somewhat larger even than the value of $\alpha=1.54$ claimed in \cite{chang2000a-a}.

\ack

A L Owczarek thanks the School of Mathematical Sciences, Queen Mary, University of London for hospitality. T Prellberg acknowledges support by EPSRC grant EP/L026708/1.

\section*{References}

\providecommand{\newblock}{}

\end{document}